\documentclass[useAMS,usenatbib]{mn2e}

\DeclareSymbolFont{cmletters}{OML}{cmm}{m}{it}
\DeclareMathSymbol{v}{\mathalpha}{cmletters}{"76}

\usepackage{amsmath}
\usepackage{amssymb}
\usepackage{epsfig}
\usepackage{graphicx}
\usepackage{ifthen}
\usepackage{latexsym}
\usepackage{rotating}
\usepackage{subfigure}
\usepackage{times,epsf}
\usepackage{txfonts}
\usepackage{varioref}
\usepackage{verbatim}
\usepackage{url}
\usepackage{color}
\usepackage[dvipsnames]{xcolor}
\usepackage[T1]{fontenc}
\usepackage{float}
\usepackage{multirow}
\usepackage{tabularx}

\newcommand{\be}{\begin{equation}}
\newcommand{\ee}{\end{equation}}
\newcommand{\bea}{\begin{eqnarray}}
\newcommand{\eea}{\end{eqnarray}}

\newcommand\aapr{Astron. \& Astrophys. Rev.}

\newcommand\apj{Astrophysical Journal}
\newcommand\apjl{Astrophysical Journal Letters}

\newcommand\apjs{Astrophysical Journal Suppl. Ser.}

\newcommand\aap{Astronomy \& Astrophysics}

\newcommand\nat{Nature}
\newcommand\prc{Physical Review C}

\newcommand\mnras{Monthly Notices of the Royal Astronomical Society}

\newcommand\physrep{Physics Reports}

\newcommand\ARAA{Ann. Rev. Astron. Astrophys.}

\newcommand\araa{\ARAA}

\newcommand{\lkhz}{$l$kHz }
\newcommand{\ukhz}{$u$kHz }
\newcommand{\FourU}{4U 1608-52}
\newcommand{\GRO}{GRO J1655-40}

  \definecolor{gray}{rgb}{0.6,0.6,0.6}
  \definecolor{green}{rgb}{0,0.6,0}

\title[Neutron Star QPOs]{Neutron Star QPOs from Oscillating, Precessing Hot, Thick Flow}
\author[P. C. Fragile]
       {P. Chris Fragile$^{1,2}$\thanks{E-mail: fragilep@cofc.edu}\\
        $^1$ Department of Physics and Astronomy, College of Charleston, Charleston, SC 29424, USA\\
        $^2$ Kavli Institute for Theoretical Physics, University of California Santa Barbara, Santa Barbara, CA 93106, USA}

\begin{document}

\maketitle

\label{firstpage}

\begin{abstract}
Across black hole (BH) and neutron star (NS) low-mass X-ray binaries (LMXBs), there appears to be some correlation between certain high- and low-frequency quasi-periodic oscillations (QPOs). In a previous paper, we showed that for BH LMXBs, this could be explained by the simultaneous oscillation and precession of a hot, thick, torus-like corona. In the current work, we extend this idea to NS LMXBs by associating the horizontal branch oscillations (HBO) with precession and the upper-kiloHertz ($u$kHz) QPO with vertical epicyclic motion. For the Atoll source 4U 1608-52, the model can match many distinct, simultaneous observations of the HBO and $u$kHz QPO by varying the inner and outer radius of the torus, while maintaining fixed values for the mass ($M_\mathrm{NS}$) and spin ($a_*$) of the neutron star. The best fit values are $M_\mathrm{NS} = 1.38 \pm 0.03 M_\odot$ and $a_* = 0.325 \pm 0.005$. By combining these constraints with the measured spin frequency, we are able to obtain an estimate for the moment of inertia of $I_\mathrm{NS} = 1.40 \pm 0.02 \times 10^{45}$ g cm$^2$, which places constraints on the equation of state. The model is unable to fit the lower-kHz QPO, but evidence suggests that QPO may be associated with the boundary layer between the accretion flow and the neutron star surface, which is not treated in this work.
\end{abstract}


\section{Introduction}
\label{sec:introduction}

Quasi-periodic oscillations (QPOs), manifestations of excess power in the Fourier transforms of X-ray light curves, were first discovered in cataclysmic variables \citep{Patterson77}. They are commonly seen in the light curves of black hole (BH) and neutron star (NS) low-mass X-ray binaries \citep[LMXBs; see reviews by][]{vanderKlis00,Remillard06}, and have recently been discovered in ultraluminous X-ray sources \citep[e.g.][]{Strohmayer03,Bachetti14} and active galactic nuclei \citep{Gierlinski08,Middleton10}. Some QPOs have frequencies suggesting they must be associated with dynamics close to the event horizon of BH systems, suggesting the possibility of using them to probe the physics of strong gravity \citep{vanderKlis06}. To realize this promise, however, a physical model for the QPOs must be found, which has, so far, been a major stumbling block. This work represents a new test of one such QPO model.

In LMXBs, QPOs are commonly divided into low- and high-frequency categories. For NS LMXBs, low-frequency (LF) QPOs have centroid frequencies in the range 0.1-60 Hz. For Z sources, identified by the shape traced out in their colour-colour diagrams \citep[CCDs;][]{Hasinger89}, LF QPOs are divided into normal branch oscillations \citep[NBOs;][]{Middleditch86}, horizontal branch oscillations \citep[HBOs;][]{vanderKlis85}, and flaring branch oscillations \citep[FBOs;][]{vanderKlis89}. Although Atoll sources do not trace out the same states in their CCDs, they nevertheless appear to exhibit corresponding HBO-like and FBO-like QPOs \citep{Motta17}. This is an important point for this paper, as the source we study is an Atoll source and one of the QPOs we focus on is the HBO-like.

NS high-frequency (HF) QPOs fall in the range 300-1300 Hz, and are thus often referred to as kiloHertz QPOs \citep{vanderKlis00}. They often occur in pairs roughly 300 Hz apart, with the higher frequency one referred to as the upper kHz QPO and the other referred to as the lower kHz QPO. These are most commonly seen in LMXBs containing low-magnetic-field NSs. The frequency of both peaks usually increases with X-ray flux, similar to HBOs.

In BH LMXBs, LF QPOs lie in the range 0.1-30 Hz and are classified as type A, B, or C \citep{Wijnands99,Casella05,Motta12}. In previous work, we showed that the type-C QPO is well explained by Lense-Thirring (LT) precession of a hot, inner flow \citep{Ingram09}. This geometric interpretation is further supported by phase-resolved studies of the type-C QPO \citep{Ingram15,Ingram16} and the prominence of type-C QPOs in high-inclination sources \citep{Homan05,Homan15}. Importantly, this model differs from the relativistic precession model \citep{Stella98}, in that the precessing object is not a test particle or orbiting blob, but an entire geometrically thick flow, such as proposed to occupy the region interior to a thin disc in the truncated disc model \citep{Esin01,Done07}. 

BH HF QPOs have frequencies $\gtrsim 100$ Hz \citep[e.g.][]{Strohmayer01,Belloni12}, usually with a lower and upper HF QPO identified in most sources. The BH HF QPOs are much weaker and harder to detect than the NS kHz QPOs \citep{Belloni12}. For this reason, some detections have required stacking observations or restricting the search to certain energy bands. This relative weakness may have to do with the lack of a boundary layer, which has been proposed to play some role in transmitting or amplifying QPO signals in NS sources \citep{Gilfanov03}. 

Recent observations of BH LMXBs \citep{Psaltis99,Motta14} have suggested that a correlation exists between certain HF and LF QPOs. Such correlations point to some common mechanism driving the QPOs themselves. In \citet{Fragile16}, we showed that a precessing, oscillating hot, thick torus, could produce frequencies that match the type-C and two HF QPOs. The HF QPOs were attributed to natural oscillations modes of the {\em same} hot, thick torus. In \citet{Fragile16}, we favored the vertical epicyclic and breathing modes, which are commonly seen in perturbed accretion flows of this type \citep{Blaes06,Mishra17,deAvellar18}. In the specific case of GRO J1655-40, this model can fit all three QPOs simultaneously using a single set of parameters.

Recent work by \citet{duBuisson19} suggests that a similar correlation exists between the kHz and HBO-like QPOs in the Atoll source \FourU. 4U 1608-52 is a fairly bright, transient NS Atoll source that has been monitored over a 16-yr period with the {\it Rossi X-ray Timing Explorer} (RXTE). It shows a rich phenomenology of fast-time variability: It contains upper- and lower-kHz QPOs, LF QPOs, likely hecto-Hz QPOs, Type I X-ray bursts, and burst oscillations. Thanks to the burst oscillations, the spin frequency of 4U 1608-52 has been measured to be 619 Hz, making it one of the most rapidly rotating accreting NSs \citep{Galloway08}. 

Since the HBO and HBO-like QPOs appear to be the NS analog of the BH type-C QPO \citep{Casella05,Motta17}, it seems reasonable that precession may also explain HBOs \citep{duBuisson19}. If so, then, by extension, our model for the BH HF QPOs may apply to the kHz QPOs of NS sources. This is what we explore in this paper.

\section{The Model}
\label{sec:model}

In \citet{Fragile16}, we presented solutions for the lowest-order oscillation modes of non-slender, hydrodynamic, non-self-gravitating, constant specific angular momentum tori around a Kerr black hole.  We use these tori as a proxy for the hot, thick flow (i.e., corona) in the truncated disc model. Geometrically, these tori are probably reasonable analogs for the real accretion flow \citep{Qian09}. However, the assumption of constant specific angular momentum is likely incorrect, as magnetohydrodynamic simulations that capture the magneto-rotational instability \citep{Balbus91,Balbus98} suggest that the inner parts of BH accretion flows have more nearly Keplerian profiles \citep{deVilliers03}. 

For NS sources, some additional considerations come into play. First, magnetic forces from the NS may impact on the accretion flow and change its structure, though QPOs are generally associated with low-magnetic field NSs \citep{vanderKlis00}. Second, for weakly magnetized NSs, the accretion flow, by necessity, must merge into a boundary layer. We expect this boundary layer, which is absent in BH sources, to quantitatively change the mode frequencies we predict. Finally, the spacetime around a NS may deviate from that of the Kerr solution \citep{Miller98}. None of these effects have been accounted for in the present work.

The second-order oscillation frequency of any mode, $i$, of a non-slender torus
\be 
  \nu_{i, m} = \left(\bar\omega_i^{(0)} + m + \beta^2 \bar\omega_i^{(2)}\right) \nu_{\rm K}
  \label{eq:nu_i}
\ee 
is composed of the slender torus (or test particle) frequency $\bar\omega_i^{(0)}/(2 \pi)$ \citep[calculated in][]{Blaes07} and the pressure correction $\bar\omega_i^{(2)}$ \citep[calculated in][]{Straub09}, where $\beta$ contains information on the thickness of the torus and $\nu_\mathrm{K} = \Omega_\mathrm{K}/(2\pi)$ is the Keplerian orbital frequency.

In this work, we focus primarily on the $m=0$ and $m=-1$ vertical epicyclic ($i=2$) modes. The $m=-1$, $i=2$ mode is the lowest order precession mode. Far from the black hole, its frequency approaches that of Lense-Thirring precession for a tilted flow. The $m=0$, $i=2$ mode manifests as global up and down oscillations of the torus about the equatorial plane of the central object.

The parameters of our model are few in number, comprised only of the mass ($M$) of the central object, the spin parameter ($a_*$), the inner ($r_\mathrm{in}$) and outer ($r_\mathrm{out}$) radii of the hot thick flow, and the polytropic index of the gas ($n$). If the mass of the central object and polytropic index can be constrained by other means, then an observation of three simultaneous QPOs is enough to constrain the three remaining free parameters (assuming all three QPOs are related to the precession/oscillation of the same structure). In the specific case of the BH LMXB \GRO, associating the type-C and two HF QPOs with LT precession and the vertical epicyclic and breathing modes, respectively, yielded $a_* = 0.63 \pm 0.12$, $r_\mathrm{in} = 6.5 \pm 0.6 r_g$, and $r_\mathrm{in} + 0.2 \le r_\mathrm{out}/r_g \le r_\mathrm{in} + 2.9$ \citep{Fragile16}, for $M_\mathrm{BH} = 5.4 \pm 0.3 M_\odot$ \citep{Beer02} and $n=3$, where $r_g = GM/c^2$ is the gravitational radius. 

For NS LMXBs, we associate the HBO (or HBO-like QPO) with precession. Doing so, we can look for what other modes match the kHz QPOs. We find that the upper-kHz QPO is well fit by vertical epicyclic motion (of the same structure that is precessing). In other words, our two modes are $\nu_\mathrm{HBO}=\nu_\mathrm{prec}=\vert \nu_{2,-1}\vert$ and $\nu_{u\mathrm{kHz}} = \nu_\mathrm{vert} = \nu_{2,0}$. We have so far not found a mode that matches the lower-kHz QPO, a point we return to in Sec. \ref{sec:discuss}. A single observation of an HBO and \ukhz QPO then allows us to constrain two of the five model parameters, though the model is obviously under-constrained at this point. However, we have previously found that our model is rather insensitive to the polytropic index, so we can safely restrict ourselves to $n=3$, leaving us with four free parameters. To proceed, one option would be to fix the mass of  the neutron star. Whenever we follow this option, we will use $M_\mathrm{NS} = 1.74 \pm 0.14 M_\odot$ \citep{Guver10}. However, we also independently find a best-fit value for the mass from our model. 

Although a single observation of two QPO frequencies is not enough to constrain our model, we can proceed by using {\em multiple} observations where the HBO and \ukhz QPO are seen together, but with differing frequencies (an option not available so far in BH sources). Since we do not expect $M_\mathrm{NS}$ nor $a_*$ to evolve significantly during a single outburst, or even from one outburst to another, a consistent model should reproduce the observed frequency pairs for {$\nu_\mathrm{HBO}$, $\nu_\mathrm{u\mathrm{kHz}}$}, with only $r_\mathrm{in}$ and $r_\mathrm{out}$ allowed to vary from one observation to the next.

\section{Application of the Model to \FourU}
\label{sec:4U}

Our model begins from a set of parameters $\{M_\mathrm{NS}, a_*, r_\mathrm{in}, r_\mathrm{out}\}$ and predicts a corresponding set of frequencies, {$\{\nu_\mathrm{prec}, \nu_\mathrm{vert}\}$ in this case. Unfortunately, we have not yet found a way to invert the calculation to start from some input frequencies and extract model parameters. Therefore, we begin our effort to match \FourU\, by constructing a grid of models, with corresponding frequency pairs $\{\nu_\mathrm{prec}, \nu_\mathrm{vert}\}$ for each one. Starting with $M_\mathrm{NS}$, we sample values from 1.3 to $2.3 M_\odot$ in steps of 0.05 $M_\odot$, consistent with physical limits based upon stellar evolution \citep{Suwa18} and realistic NS equations of state \citep[EOS;][]{Akmal98}. The dimensionless spin of a neutron star is theoretically constrained to be $a_* \lesssim 0.7$ \citep{Miller98}, but observationally seems to be somewhat smaller, $a_* \le 0.3$ \citep{Miller15}. However, since 4U 1608-52 is one of the fastest rotating neutron stars, with an observed spin period (from burst oscillations) of 619 Hz \citep{Galloway08}, we consider spins in the range $0.05 \le a_* \le 0.45$, in steps of 0.05. For the inner radius, we do not expect it to lie inside the innermost stable circular orbit, so we set this as our lower value, $r_\mathrm{in} \ge r_\mathrm{isco}$. However, the accretion flow could truncate further out, especially in the case of tilted flows, because of additional angular momentum extraction due to standing shocks associated with the tilt \citep{Fragile09,Dexter11,Generozov14} or in the case of NSs because of a magnetospheric radius $r_\mathrm{m} > r_\mathrm{isco}$. While the upper value is not formally restricted, we limit ourselves here to $r_\mathrm{in} \le 10 r_g$, in steps of $0.5r_g$. For the outer radius, there is no formal restriction, other than the obvious one that it must lie outside $r_\mathrm{in}$. In this work, we set $r_\mathrm{in} + 0.2 \le r_\mathrm{out}/r_g \le 30$. However, as noted in \citet{Fragile16}, some combinations of $r_\mathrm{in}$ and $r_\mathrm{out}$ fail to produce a valid torus solution, given the constraints of our analytic model, and so, produce no corresponding frequency pair.

With these parameter choices, we are able to produce 35,052 model frequency pairs to compare to simultaneous observations of HBO and \ukhz QPOs. For \FourU, we restrict ourselves to the 8 unambiguous "confirmed triplets" and the 8 "tentative triplets" for which the \ukhz Q-factor could be properly fit, as identified in Table A1 of \citet{duBuisson19}.  Although we are not trying to match the lower-kHz QPO with our model, we stick to the confirmed and tentative triplets, which include the HBO-like, \lkhz and \ukhz QPOs, as these provide more reliable identifications of $\nu_\mathrm{HBO}$ and $\nu_{u\mathrm{kHz}}$. 

Nominally, in order to estimate the likelihood that any one of our models fits the data, we would calculate the joint probability for each combination of $M_\mathrm{NS}$ and $a_*$ as 
 \begin{equation}
-\ln [ P (M_\mathrm{NS}, a_*)] =  \sum_i \frac{[\nu_{\mathrm{prec}, i} - \nu_{\mathrm{HBO}, i}]^2 + [\nu_{\mathrm{vert},i} - \nu_{u\mathrm{kHz}, i}]^2 }{ (\sigma_{\mathrm{HBO}, i} \cos \Phi_{i})^2 + (\sigma_{u\mathrm{kHz}, i}\sin \Phi_{i})^2 } ~,
\label{eqn:prob}
\end{equation}
where 
\begin{equation}
\Phi_{i} = \arctan\left(\frac{\nu_{u\mathrm{kHz},i} - \nu_{\mathrm{vert},i}}{\nu_{\mathrm{HBO},i} - \nu_{\mathrm{prec},i}}\right) ~,
\end{equation}
$\sigma$ is the geometric mean of the asymmetric error values found in the observed data, and $\nu_{\mathrm{prec},i} = \nu_\mathrm{prec}(M_\mathrm{NS}, a_*, r_\mathrm{in}, r_\mathrm{out})$ and $\nu_{\mathrm{vert},i} = \nu_\mathrm{vert}(M_\mathrm{NS}, a_*, r_\mathrm{in}, r_\mathrm{out})$ represent the model frequency pair that lies closest to observation $i$, as defined by minimizing the error in eq. (\ref{eqn:prob}). However, in practice, we find this formal probability to be quite low, $P(M_\mathrm{NS}, a_*) \lesssim 10^{-6}$, when applied in this manner. This is due to the relatively sparse sampling of our model data and the generally small uncertainties (relative to the overall range of values) ascribed to the observed QPO frequencies. This makes some of the error terms contained in the sum in eq. (\ref{eqn:prob}), quite large, driving the overall probability down. Yet, in some cases, as we will show, the observed frequency pairs lie entirely within the range covered by our model. In such cases, if we had perfect sampling of our model space, the corresponding error terms $(\nu_\mathrm{model} - \nu_\mathrm{obs})^2$ would all be zero and the probability that such a model matches the data should be 1. Therefore, we modify the sum in eq. (\ref{eqn:prob}) by excluding any observations that lie within the model space (defined as any observation whose error bars overlap the model space). Figure \ref{fig:massvspin} shows the resulting distribution of $\ln[P(M_\mathrm{NS}, a_*)]$. From this, we see that our successful models all cluster around a typical neutron star mass ($M_\mathrm{NS} \approx 1.4 M_\odot$) with a moderately high spin parameter ($a_* = 0.3$-0.35). With the sparse sampling of our model parameter space, it is difficult to get well fit error estimates for our mass and spin, but formally, the best-fit values and $1\sigma$ errors are $M_\mathrm{NS} = 1.38 \pm 0.03 M_\odot$ and $a_* = 0.325 \pm 0.005$.

\begin{figure}
\includegraphics[width = 0.48\textwidth]{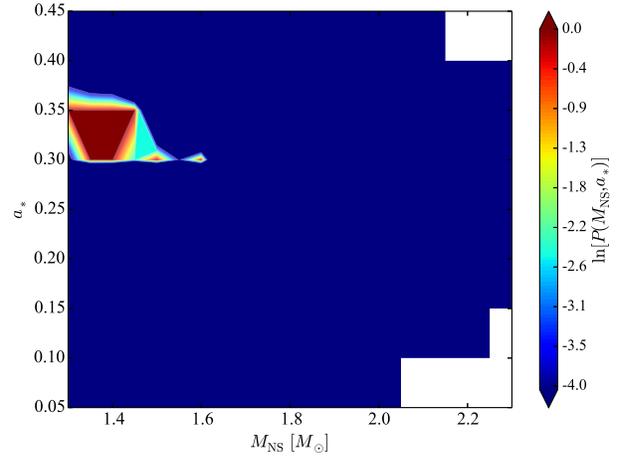}
\caption{Contour plot showing the joint probability from eq. (\ref{eqn:prob}) associated with each combination of neutron star mass and spin. Only a relatively small area of parameter space gives reasonable likelihoods.}
\label{fig:massvspin}
\end{figure}

To better illustrate how our model frequency pairs span the parameter space and correlate with the observed frequency pairs, we plot one of our successful models ($M_\mathrm{NS} = 1.35 M_\odot$, $a_*$ = 0.35) in Figure \ref{fig:nuVnu_1.35}. The blue and green symbols show the confirmed and tentative frequency pairs from \citet{duBuisson19} that we are trying to match. The gray and red symbols show model frequency pairs corresponding to particular values of $r_\mathrm{in}$ and $r_\mathrm{out}$, with the red diamonds indicating designated matches. This illustrates a case where all of the observed frequency pairs lie within the parameter space defined by the model points. Although some of our declared matches lie outside the formal error bars of their corresponding observations, the probability of a match is still considered high because there is some combination of $r_\mathrm{in}$ and $r_\mathrm{out}$ that {\em could} produce a perfect match for each observation in this case. It is important to note that the scale on the two axes is different -- a distance of 10 Hz in the horizontal direction appears much larger in the plot than a distance of 10 Hz in the vertical direction. This explains why model points that appear to lie a long way away from an observation in the horizontal direction will be chosen as designated matches over model points that appear to lie closer in the vertical direction. In all cases, each designated match is chosen as the model point that contributes the smallest amount to the sum in eq. (\ref{eqn:prob}).

\begin{figure}
\includegraphics[width = 0.48\textwidth]{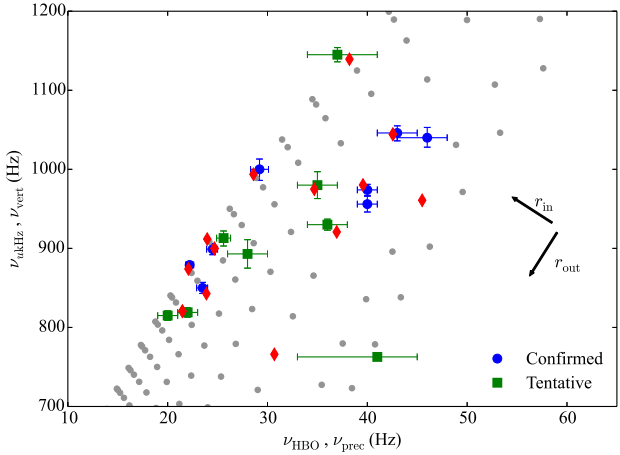}
\caption{Frequency-frequency plot comparing our sampling of model pairs (gray circles and red diamonds) for $M_\mathrm{NS} = 1.35 M_\odot$ and $a_* = 0.35$ to the confirmed and tentative triplets of \citet{duBuisson19}. Black arrows show roughly how the model frequencies vary with changes in $r_\mathrm{in}$ and $r_\mathrm{out}$. The red diamonds denote declared matches, as summarized in Table \ref{tab:matches}. Note that the scale on the two axes are different, so small displacements in the vertical direction contribute much larger errors, generally, than similar length displacements in the horizontal direction.}
\label{fig:nuVnu_1.35}
\end{figure}

Each unique set of $r_\mathrm{in}$ and $r_\mathrm{out}$ corresponding to a matching frequency pair in Figure \ref{fig:nuVnu_1.35} is reported in Table \ref{tab:matches} to give an idea of their range. The values of $r_\mathrm{in}$ are consistent with numerical studies of tilted, hot thick accretion flows \citep{Fragile09, Dexter11}, as appropriate for our precession picture \citep{Ingram09} or modestly magnetized NSs \citep[$B \lesssim 10^8$ G;][]{Davidson73}. The outer radius should roughly correspond to the truncation radius of the cold, geometrically thin, optically thick disc that feeds matter into the neutron star. It may be possible, therefore, to independently verify the plausibility of our model by measuring this truncation radius, through perhaps Fe-line reflection modeling \citep[e.g.,][]{Cackett10}, and confirming that it agrees with $r_\mathrm{out}$. 

\begin{table}
\begin{center}
\begin{tabular}{ccccc}
\hline
 $M_\mathrm{NS}$, $a_*$ & $r_\mathrm{in}$ &  $r_\mathrm{out}$ & $\nu_\mathrm{vert}$ & $\nu_\mathrm{prec}$ \\
 & $(r_g)$ &  $(r_g)$ & (Hz) & (Hz) \\
\hline
\multirow{14}{*}{$1.35 M_\odot$, 0.35}  & 5.0 & 9.7 & 961 & 45.5 \\
& 5.5 & 9.2 & 1044 & 42.6 \\
& 5.5 & 9.7 & 980 & 39.6 \\
& 5.5 & 10.2 & 921 & 36.9 \\
& 5.5 & 11.7 & 766 & 30.7 \\
& 6.0 & 9.7 & 975 & 34.7 \\
& 7.0 & 7.7 & 1139 & 38.2 \\
& 7.0 & 10.7 & 843 & 23.9 \\
& 7.5 & 9.7 & 900 & 24.7 \\
& 7.5 & 10.7 & 821 & 21.5 \\
& 8.0 & 8.2 & 994 & 28.6 \\
& 8.5 & 8.7 & 912 & 24.0 \\
& 8.5 & 9.2 & 874 & 22.1 \\
\hline
\multirow{13}{*}{$1.6 M_\odot$, 0.3} & 5.0 & 8.7 & 925 & 38.6 \\
& 5.5 & 7.2 & 1150 & 44.1 \\
& 5.5 & 8.7 & 937 & 33.6 \\
& 5.5 & 10.2 & 771 & 26.9 \\
& 6.0 & 7.7 & 1047 & 35.3 \\
& 6.0 & 8.2 & 984 & 32.2 \\
& 6.0 & 9.7 & 821 & 25.4 \\
& 6.5 & 7.2 & 1069 & 34.6 \\
& 6.5 & 7.7 & 1010 & 31.3 \\
& 6.5 & 8.7 & 902 & 26.3 \\
& 6.5 & 9.2 & 853 & 24.2 \\
& 6.5 & 9.7 & 808 & 22.5 \\
& 7.0 & 8.2 & 918 & 25.6 \\
& 7.5 & 8.2 & 881 & 23.1 \\
\hline
\end{tabular}
\caption{A table of each unique matching \{$r_\mathrm{in}$, $r_\mathrm{out}$\} pair for the given combinations of $M_\mathrm{NS}$ and $a_*$.
\label{tab:matches}}
\end{center}
\end{table}

If, instead of allowing $M_\mathrm{NS}$ to be a free parameter, we require it to be fixed at $M_\mathrm{NS} = 1.74 \pm 0.14 M_\odot$, based on the observation of \citet{Guver10}, then we see that the only acceptable fit in Figure \ref{fig:massvspin} occurs at $M_\mathrm{NS} = 1.6 M_\odot$ and $a_* = 0.3$. Figure \ref{fig:nuVnu_1.6} illustrates how the model frequency pairs (gray circles) and matches (red diamonds) are distributed in this case. The main difference between Figures \ref{fig:nuVnu_1.35} and \ref{fig:nuVnu_1.6} is that in Figure \ref{fig:nuVnu_1.6}, the model covers a narrower range in the horizontal ($\nu_\mathrm{HBO}$, $\nu_\mathrm{prec}$) direction, making it harder to match observations near the extremes. Table \ref{tab:matches} again presents the values of $r_\mathrm{in}$ and $r_\mathrm{out}$ corresponding to the declared matches in Figure \ref{fig:nuVnu_1.6}.

\begin{figure}
\includegraphics[width = 0.48\textwidth]{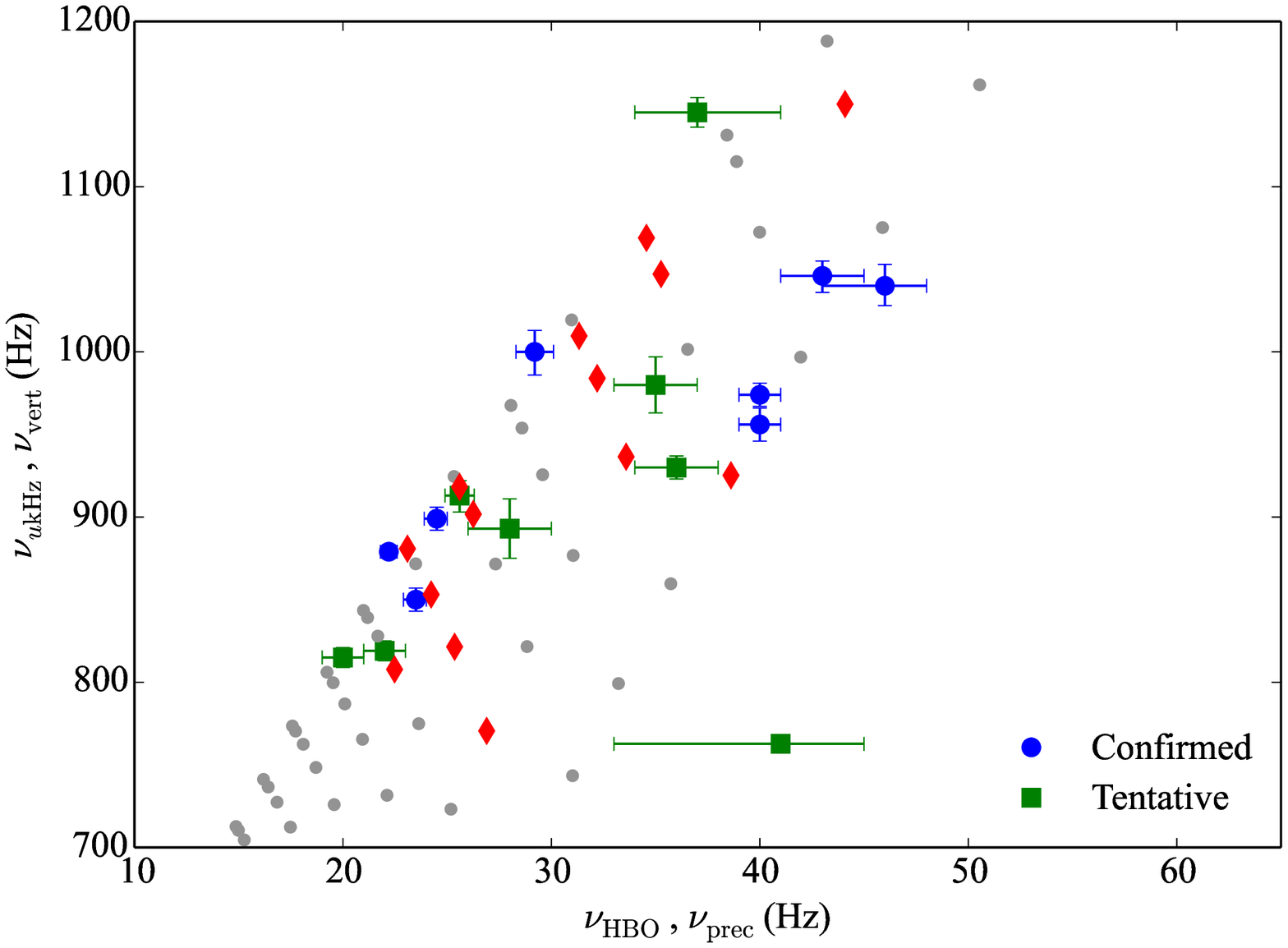}
\caption{Same as Fig. \ref{fig:nuVnu_1.35}, but for $M_\mathrm{NS} = 1.6 M_\odot$ and $a_* = 0.3$.}
\label{fig:nuVnu_1.6}
\end{figure}

The trend of narrowing of the model parameter space continues with increasing mass, thus explaining the strong exclusion of values $M_\mathrm{NS} > 1.6 M_\odot$. Our model also strongly disfavors all low-spin ($a_* \le 0.25$) combinations, because they cannot produce precession frequencies high enough to match the observed HBO frequencies in \FourU.

\section{Constraining the NS EOS}
\label{sec:EOS}

Using the mass and spin derived from our model, along with the observed spin frequency of \FourU, $\nu_\mathrm{spin} = 619$ Hz, we can estimate the moment of inertia of the neutron star \citep[e.g.,][]{Middleton18}
\begin{equation}
I_\mathrm{NS}  = \frac{a_*GM_\mathrm{NS}^2}{2\pi c \nu_\mathrm{spin}} = (1.40 \pm 0.02) \times 10^{45} \mathrm{~g~cm}^2
\label{eqn:inertia}
\end{equation}
or $I_\mathrm{NS}/M^{3/2} = 43.4 \pm 0.6$ km$^2$ $M_\odot^{-1/2}$. Based upon Figure 1 of \citet{Lattimer05}, this puts the neutron star in \FourU\, closest to the PCL2, and between the AP4 and ENG, EOS \citep[see][for a description of each of these models]{Lattimer01}. Note that eq. (\ref{eqn:inertia}) assumes $a_* \equiv Jc/GM^2 = I\omega c/GM^2 = 2\pi I \nu_\mathrm{spin} c/GM^2$, which is only valid if the star is uniformly rotating (solid-body rotation). We have also assumed that the spacetime exterior to a neutron star is well described by the Kerr metric. A better estimate of the moment of inertia would require one to choose a specific equation of state, solve for the spacetime metric self-consistently, and then re-solve for the QPO frequencies in this new spacetime.

We can also use our value for $I_\mathrm{NS}$ to get rough estimates for the neutron star radius by assuming two simple models for its moment of inertia: those of a hollow
\begin{equation}
I_\mathrm{hollow} = \frac{2}{3} M r^2
\end{equation} 
and a solid
\begin{equation}
I_\mathrm{solid} = \frac{2}{5} M r^2
\end{equation} 
sphere. We find values of  $r_\mathrm{hollow} = 8.7$ km and $r_\mathrm{solid} = 11.3$ km, respectively, consistent with the current best estimates for neutron star radii \citep[e.g.,][]{Lattimer07, Demorest10}.

These values for the moment of inertia and radius of the NS should not be taken too seriously, however. We are not claiming to have directly measured or carefully modeled either quantity. Nevertheless, it is a nice consistency check on the QPO model that it gives reasonable parameters.

\section{Discussion and Conclusions}
\label{sec:discuss}

In this paper, we presented a model to explain the HBO and upper-kHz QPO of NS LMXBs as the precession and vertical epicyclic oscillation of a hot, geometrically thick accretion flow filling the region interior to a truncated, thin accretion disc. For the case of 4U 1608-52, this model was able to successfully match all of the confirmed and well-characterized, tentative simultaneous observations of these two QPOs with reasonable values for the mass ($M_\mathrm{NS} = 1.38 \pm 0.03 M_\odot$) and spin ($a_* = 0.325 \pm 0.005$) by only varying the inner and outer radius of the thick accretion flow.

The model tightly constrains the possible parameter space. Spins $a_* \le 0.25$ are strongly disfavored, as they are not able to produce high enough precession frequencies to match the HBO. Likewise, high mass neutron stars are disfavored because they cannot produce a wide enough range of precession frequencies. These constraints lead to pretty tight limits on the moment of inertia of the neutron star in \FourU, as derived from our model, $I_\mathrm{NS} = (1.40 \pm 0.02) \times 10^{45}$ g cm$^2$.

However, the limits quoted in this work do not account for systematic effects likely to impact this model. The biggest may come from our use of a relatively simple, analytic model for the hot, thick flow and its associated oscillation frequencies. Yet, this is a complicated geometry we are picturing -- a possible boundary layer interacting with a hot, thick flow, interacting with a cold, thin disc. Thus, the true frequencies, and hence the derived parameters, are probably somewhat different. Nevertheless, the fact that the frequency range covered by our model agrees so well with the observed frequency range in \FourU\, makes this a promising model, worthy of further study.

There are, of course, many caveats. One issue is that, in our analogous model for BH QPOs \citep{Fragile16}, there was a third oscillation frequency present -- the breathing mode. The breathing mode has an even higher frequency than the vertical epicyclic one, so, if our current analysis is correct and the \ukhz QPO in NSs is associated with the vertical epicyclic mode, then a straightforward application of our BH model to NSs would predict another QPO (associated with the breathing mode) at $\gtrsim 2000$ Hz. This is below the Nyquist frequency for RXTE, so if a strong QPO were present at this frequency, it should have been observed. Thus, one must conclude that: 1) something in the NS case suppresses this mode; 2) we have misidentified the modes in either the BH or NS systems; or 3) the breathing mode only shows up under special circumstances not met in \FourU. The last possibility is certainly plausible, as the upper HFQPO is only rarely, and weakly, seen in BH LMXBs, and only once was it observed simultaneously with the lower HF QPO \citep{Motta14}.

The model also relies on the truncated disk geometry to explain the evolution of the QPO frequencies during the outburst. However, iron-line reflection \citep{Miller06} and reverberation mapping \citep{Kara19} studies have suggested that accretion disks are either not truncated or truncated at small, {\em constant} radii. If the disk is still truncated, but at a constant radius, then our QPO model may still be able to work if there were another parameter, such as disk surface density, that would vary appropriately over the course of the outburst to explain the evolution of the QPO frequencies. This is something we are currently exploring. If, on the other hand, disks are truly not truncated, even in states associated with QPOs, then it may be that QPOs are still associated with oscillation modes of the corona, but it becomes harder to see how precession could be one of those modes.

Perhaps the biggest issue, though, is that our model does not explain the \lkhz QPO. The trouble is, this QPO does not appear to have the correct frequency or scaling relative to the other QPOs to match any of the torus oscillation modes identified in \citet{Fragile16}. On the one hand, this appears to contradict decades of evidence that the two kHz QPOs are correlated \citep{Psaltis99,Stella99}. On the other hand, there is growing evidence that the \lkhz QPO is not associated with the same part of the accretion flow as the \ukhz: 1) the two QPOs follow different tracks in a quality factor versus frequency diagram \citep{Barret05}; and 2) the spectral-timing behavior of these two QPOs is systematically different \citep{Peille15,Troyer18}. In particular, the correlation of the \lkhz QPO with the mass accretion rate and NS magnetic field strength suggest it may be associated with the magnetospheric radius and boundary layer \citep{Erkut16}. This QPO also exhibits energy-dependent phase lags that are consistent with a radiation-pressure-supported boundary layer (O. Blaes, private communication). We are proposing, then, that this QPO may be driven by oscillations, not of the corona, but of the boundary layer itself, which are not covered in our current model. A future goal of our work will be to study the properties of such oscillations and try to connect them to those of the hot accretion flow to create a single model that can explain the HBO, the \ukhz QPO, {\em and} the \lkhz QPO.

\section{Acknowledgements}

Special thanks go to Erika Hamilton and Josh White for their help with this project. Additional thanks go to Omer Blaes, Jason Dexter, and Sara Motta for helpful discussions. This work was supported by National Science Foundation grants AST-1616185, AST-1907850, and PHY-1748958.
 
\bibliographystyle{mn2e}

\end{document}